# Physics Results from the AMANDA-B10 Neutrino Telescope

A. Hallgren for the AMANDA Collaboration

Department of High Energy Physics, Uppsala University, Box 535, S-751 21Uppsala, Sweden

The data from the first year of operation of the 10-string, AMANDA-B10, high-energy neutrino detector array at the South Pole have been analyzed and searched for evidence of neutrinos from cosmic sources. Differently optimized selection criteria have been used and limits on the flux of high-energy neutrinos from a variety of objects have been derived.

## INTRODUCTION

High-energy neutrinos from various types of cosmic sources are predicted by many models. Measurement of the flux and energy distribution of these neutrinos is important for the understanding of processes in our universe, in particular the regions with intense energy transformations, and will shed light on the validity of the models. The experimental task is however very challenging. This contribution reports results obtained by the AMANDA collaboration on the neutrino flux. An overview of the field of High Energy Neutrino Astronomy is given elsewhere in this conference [1].

## THE AMANDA-B10 NEUTRINO DETECTOR

The AMANDA Collaboration has built a high-energy neutrino telescope situated in the Antarctic glacier at the geographic South Pole. The detector consists of an array of Optical Modules (OM), each containing an 8-inch photo-multiplier, that detects Cherenkov light from muons and electrons induced by neutrino interactions in the surrounding medium. The OMs are distributed along cable strings in the ice. The first 4 strings, containing 86 OMs, were installed in the austral summer 1995-1996 and the results can be found in [2]. In the following austral summer the detector was augmented with 6 additional strings and a total of 302 installed OMs was reached, forming the AMANDA–B10 detector. The instrumented volume of ice is roughly cylindrical with a diameter of 120 m and situated between 1500 and 2000 m depth. The detector has since then been further augmented and presently contains 677 OMs distributed over 19 strings. The installation of the extended detector was completed in January 2000. A figure showing the present detector can be found in [3]. A further increase to kilometer-scale is planned within the IceCube project [4].

The optical properties of the ice at the depth of the detector have been measured by means of artificial light sources installed in the ice. The optical properties have been found to be depth dependent, correlated to variations in dust concentration [5]. Typical values for the absorption and scattering lengths are of order 100 m and 25 m respectively.

## DATA SAMPLE AND ANALYSIS STRATEGY

The analysis reported here is based on $10^9$ events taken with the AMANDA–B10 array during the austral winter period from April to November 1997. The trigger required $\geq 16$ OM signals arriving to the surface electronics within 2.5 $\mu$s, yielding an event rate close to 100 Hz. The integrated live-time corresponds to 130 days.

The overwhelming majority of the events is due to muons from cosmic rays entering the detector from above. To avoid this important background only upmoving events are considered as neutrino candidates. Most of the events were thus removed by a zenith angle cut on the result of a fast track fit. In the subsequent maximum likelihood reconstruction a more realistic photon arrival time distribution, mainly accounting for scattering in the ice, was used. Fits were performed both to the muon track and to the electron shower hypothesis. Quality indicators were then derived and used for further

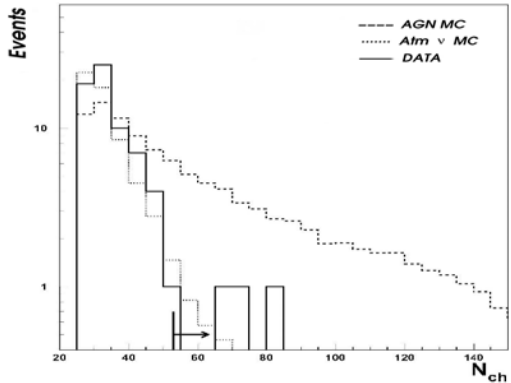

Figure 1. OM multiplicity distributions for experimental data and for simulated atmospheric neutrinos and neutrinos from AGNs. The simulated AGN neutrinos have an $E^{-2}$ spectrum and a flux corresponding to $10^{-5}$ GeV cm$^{-2}$ s$^{-1}$ sr$^{-1}$. The cut used for the limit on the diffuse flux is shown.

selection, e.g. track length, number of unscattered hits, hit distribution along the track etc.

The efficiency and the background rejection power have been verified by using the known flux of neutrinos produced in the atmosphere. These results are reported at this conference [3]. The correspondence between Monte Carlo (MC) simulations and experimental data was verified by this analysis, thus establishing AMANDA as an operational high-energy neutrino telescope. The selection criteria used in searches for neutrinos from other types of sources have been optimized to account for differences in energy distributions and in background, as briefly described below.

### DIFFUSE FLUX OF HIGH-ENERGY $\nu_\mu$

A "diffuse" flux of high-energy neutrinos is predicted as the sum of fluxes from a large number of sources distributed over the sky. These neutrinos have a harder spectrum than the atmospheric neutrinos. A selection on neutrino energy can thus be made to enhance neutrinos from the cosmic sources. The correlation between the neutrino energy and the OM multiplicity in AMANDA, for quality-selected events, was studied with MC simulations. The difference in the multiplicity distributions is shown in Fig. 1. An optimal value for a cut in the multiplicity at 53 OMs was derived from the MC study. The corresponding energy band is about 20 – 800 TeV (range of half peak sensitivity). In the preliminary analysis three events were observed above this multiplicity cut, consistent with the background of atmospheric neutrinos. The corresponding, preliminary, 90% confidence level upper limit on the neutrino flux, $E^2\Phi\nu$ is $0.9*10^{-6}$ GeV cm$^{-2}$s$^{-1}$sr$^{-1}$. For details see [6]. A comparison with other experimental limits and with theoretical bounds can be found in these proceedings [1].

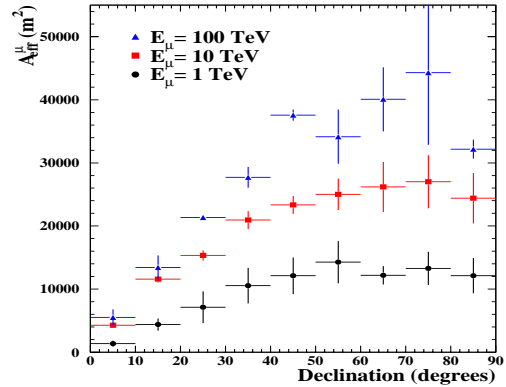

Figure 2. Effective area for muons versus zenith angle and for different muon energies.

### HIGH-ENERGY $\nu_\mu$ FROM POINT SOURCES

Another analysis searched for an excess of neutrinos from restricted angular regions in the sky, i.e. from point sources. Known astrophysical objects being candidate sources have been examined in addition to an unbiased sky search. Benefiting from the angular restriction, the point source analysis [7] uses less strict quality criteria to optimize the signal sensitivity. The background is strongly reduced by the spatial correlation condition. The angular resolution of ~3.5° was determined from MC and verified using down-going muon events coincident with shower observations in the SPASE surface array. The final selection criteria were determined from MC studies. The effective area obtained varies with energy and with zenith angle, reaching about 30000 m$^2$ for 100 TeV muons, fig 2. A selection of ~ 800 events was obtained. No clustering was observed, yielding a preliminary limit of $E^2\Phi\nu \approx 10^{-6}$ GeV cm$^{-2}$s$^{-1}$sr$^{-1}$ on the flux from point sources, see fig 3.

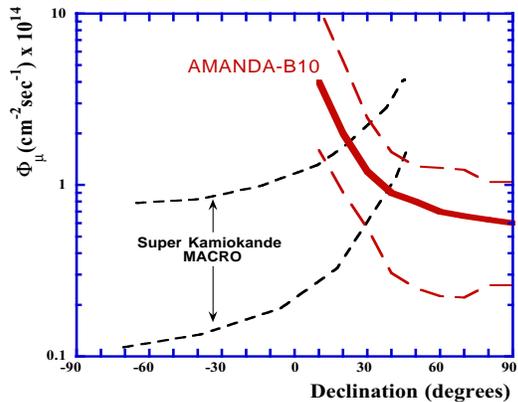

Figure 3. Upper limits (90% CL) for muon flux induced by neutrinos from point sources. For AMANDA an unbiased "all sky" preliminary limit is shown and the range indicated corresponds to uncertainties in the calculation of the effective area. For MACRO and Super Kamiokande the range of limits for a sample of known objects is shown [9].

## $\nu_\mu$ FROM GAMMA RAY BURSTS

In the search for high-energy neutrinos produced by Gamma Ray Bursts (GRB) a sample of 78 BATSE registered events were investigated. The additional criterion of time coincidence allows for a further relaxation of the quality selection criteria, giving increased sensitivity. The cuts were adapted to a predicted energy distribution [8]. The background in the detector was monitored during one hour before and one hour after the BATSE recorded GRB time. The time interval recorded by BATSE for the duration of the burst was used for signal extraction. Variations in GRB parameters and distance predict large individual GRB differences in the neutrino flux. No burst gave a statistically significant excess. A combined upper limit on the neutrino flux was obtained, $\Phi_\nu^{90\%}$=1 (5) (0.2)·$10^{-10}$ cm$^{-2}$s$^{-1}$ for boost factors $\Gamma$ = 100 (300) (1000).

## SEARCH FOR $\nu_e$ CASCADES

The light pattern caused by electrons from a high-energy $\nu_e$ interaction is better fitted to a spherical model than to the muon hypothesis, and can thus be identified. The spatial resolution was determined with simulation of AMANDA-B10 to be 4-5 m for the vertex position. The energy resolution was found to be better than 0.45 in log$_{10}$E. After elimination of the background no events remained in the sample, yielding a 90% CL upper limit on the $\nu_e$ flux $E^2\Phi_{\nu e}\leq$ 7·$10^{-6}$ GeV cm$^{-2}$ s$^{-1}$ for 10 < $E_\nu$ < 50 TeV.

## NEUTRALINO AND MAGNETIC MONOPOLE

Neutralinos ($\chi$), being relic supersymmetric particles from the Big Bang, are candidates for the dark matter in the universe. Neutralinos would accumulate in the Earth core where they would annihilate and produce $\nu_\mu$ visible in the AMANDA-B10 detector. The absence of an excess of muon events in the vertical direction provides a preliminary limit on a $\chi$ induced muon flux $\Phi_\mu$ = 2·$10^{-3}$ km$^{-2}$yr$^{-1}$ for a $\chi$ mass around 1 TeV.

Relativistic magnetic monopoles traversing the detector would give rise to a large quantity of Cherenkov light. The data has been searched for events with very high multiplicity. Their absence yields a preliminary limit of 0.6·$10^{-16}$ cm$^{-2}$s$^{-1}$sr$^{-1}$ for the flux of magnetic monopoles with $\beta$=1.

## CONCLUSIONS

Data from the neutrino telescope AMANDA-B10 collected in 1997 have been analyzed for presence of high-energy $\nu_\mu$ and $\nu_e$ events from cosmic sources, for dark matter candidates and for magnetic monopoles. The simulation of the detector shows good agreement with the experimental data and the severe background conditions have been mastered. The limits obtained are comparable to limits achieved by other detectors with much longer exposure. Analysis of data taken with AMANDA-B10 in the years 1998 and 1999 as well as later data from the expanded AMANDA-II array [3] will further improve the sensitivity to high-energy neutrinos from extra-galactic sources.